\documentclass[pra, twocolumn, 
 amsmath,amssymb,
 aps,
 hyperref,
]{revtex4}

\usepackage{graphicx}
\usepackage{dcolumn}
\usepackage{bm}
\usepackage{physics}
\usepackage{tabularx}
\usepackage{mathrsfs}
\usepackage{verbatim}
\newcolumntype{Y}{>{\centering\arraybackslash}X}
\usepackage[
colorlinks=true,
linkcolor=blue,
anchorcolor=blue,
citecolor=blue,
urlcolor=blue,
]{hyperref}

\begin{document}

\preprint{APS/123-QED}

\title{Pair-Condensation Peak in Radio-Frequency Dimer-Projection Spectroscopy of a Unitary Fermi Gas}

\author{Zhixiu Peng}
\author{Zhenhua Yu}%
 \email{yuzhh5@mail.sysu.edu.cn}
\affiliation{%
 Guangdong Provincial Key Laboratory of Quantum Metrology and Sensing, School of Physics and Astronomy\\
 Sun Yat-Sen University (Zhuhai Campus), Zhuhai, Guangdong 519082, China\\
 and State Key Laboratory of Optoelectronic Materials and Technologies, Sun Yat-Sen University
 (Guangzhou Campus), Guangzhou, Guangdong 510275, China
}%




\date{\today}

\begin{abstract}
Radio-frequency dimer-projection spectroscopy has recently emerged as a powerful probe of strongly interacting Fermi gases, yet its frequency-resolved spectrum remains largely unexplored. Here we show that the spectrum exhibits two qualitatively distinct features: a broad quasiparticle continuum governed by fermionic pairing and a sharp pair-condensation peak arising from the coherent projection of condensed Cooper pairs onto the final-state dimer. To reveal this physics, we develop a point-dimer approximation that exploits the separation between the molecular and many-body length scales and expresses the radio-frequency response directly in terms of pair correlations in the initial many-body state. Within the BCS mean-field theory, we derive the frequency-dependent spectrum, identify the microscopic transfer processes underlying its structure, and show that the continuum possesses a threshold determined by the superfluid gap. Our theory establishes radio-frequency dimer-projection spectroscopy as a unified probe of short-range correlations, fermionic pairing, and long-range coherence in strongly interacting Fermi gases.
\end{abstract}

\maketitle


\emph{\textcolor{blue}{Introduction.---}}
Radio-frequency (RF) spectroscopy is one of the most versatile probes of strongly interacting ultracold atomic gases. Owing to its ability to transfer atoms between internal hyperfine states while imparting negligible momentum, RF spectroscopy provides direct access to interaction effects, pairing correlations, quasiparticle excitations, and universal properties of quantum many-body systems. Since its first application to degenerate Fermi gases by Gupta \emph{et al.}~\cite{Gupta2003}, RF spectroscopy has played a central role in the study of the BEC--BCS crossover. It has revealed pairing phenomena~\cite{Chin2004,Schunck2007,Schirotzek2008,Mukherjee2019}, quasiparticle excitations~\cite{Schirotzek2008}, and single-particle spectral functions through momentum-resolved measurements~\cite{Stewart2008}. More generally, RF spectroscopy has become a fundamental tool for investigating strongly interacting quantum gases, analogous to photoemission spectroscopy in condensed-matter systems~\cite{Vale2021}.

\begin{figure}[t]
\includegraphics[width=1.1\columnwidth]{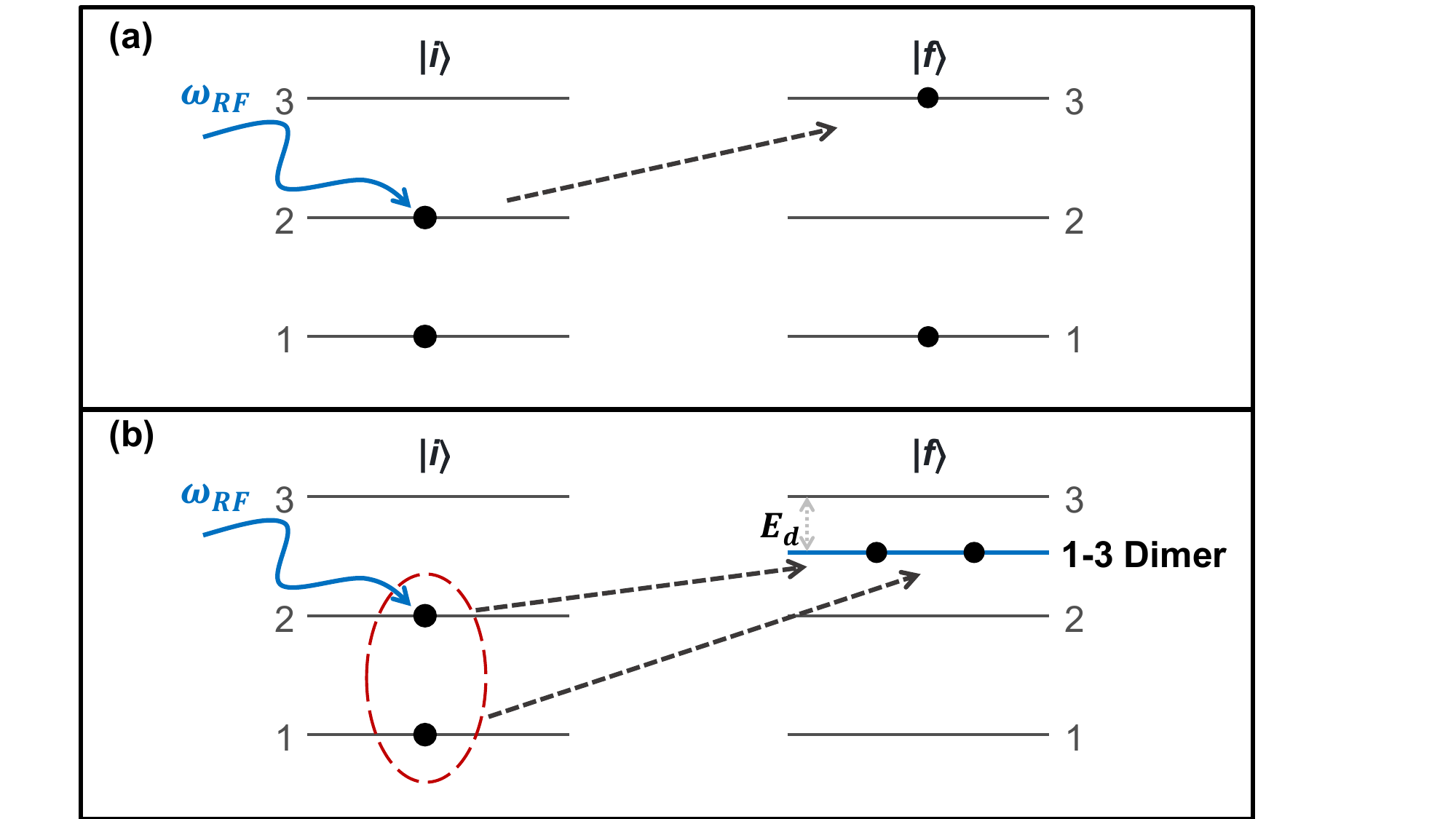}
\caption{\label{schematic} Schematic of the radio-frequency spectroscopy. Starting with interacting atoms only occupying the internal states $\ket1$ and $\ket2$, (a) most previous experiments tune the excitation frequency $\omega_{\rm RF}$ near resonant with or blue-detuned with respect to the atomic transition frequency between states $\ket2$ and $\ket3$. (b) A recent experiment~\cite{Xie2026} red-detunes $\omega_{\rm RF}$ and projects an interacting atom pair onto a medium-bound dimer supported by the final-state interaction channel.}
\end{figure}

Most previous RF experiments probe transitions from an interacting initial state to a weakly interacting continuum final state. Consequently, the measured response is dominated either by near-resonant atomic excitations or by the high-frequency continuum [see Fig.~\ref{schematic}(a)], from which quantities such as the pairing gap or Tan's contact are extracted. The high-frequency tail of the RF spectrum was shown theoretically to obey a universal $\omega^{-3/2}$ scaling governed by the contact~\cite{Tan2008a,Tan2008b,Tan2008c,Braaten2010,Schneider2010}, and has subsequently been employed experimentally to measure short-range correlations in strongly interacting Fermi gases~\cite{Stewart2010,Sagi2012,Shkedrov2018}. Exact sum rules further relate the first moment of the RF spectrum to interaction-induced clock shifts and many-body correlation functions~\cite{Baym2007,Punk2007,Zhang2008,Zhang2009,Langmack2012}. These developments have established RF spectroscopy as a unique probe connecting microscopic pair correlations with macroscopic thermodynamic properties. Beyond equilibrium properties, Tan's contact has also
been proposed as a sensitive probe of nonequilibrium
many-body dynamics, including equilibration,
hydrodynamic attractors, and interaction quenches~\cite{Fujii18,Nishida19,Enss19,Hofmann20,Fujii24,Mazeliauskas26,Son10,Huang19,Natu13,Corson15,Heras19}.

A different spectroscopic scheme has recently been demonstrated by Xie \emph{et al.}~\cite{Xie2026}. Instead of transferring atoms into a continuum state, the RF field projects an interacting atom pair onto a medium-bound dimer supported by the final-state interaction channel [see Fig.~\ref{schematic}(b)]. Molecular conversion spectroscopy has a long history in ultracold gases through photoassociation and photoexcitation experiments, although those techniques probe different transition channels than RF dimer projection does~\cite{Partridge05,Werner09,Wang21,Paintner19,Liu22,Jager24_1,Jager24_2,Journeaux26}. Owing to the large separation between the dimer binding energy and the many-body energy scale, the dimer resonance forms a well-isolated spectral feature whose integrated weight is directly proportional to Tan's contact. Furthermore, because the spectral weight is concentrated into a discrete resonance rather than distributed over a broad high-frequency continuum, the contact can be measured on timescales shorter than the Fermi time. 
This capability therefore complements conventional RF spectroscopy and provides access to previously inaccessible dynamical regimes.


Despite this experimental advance, the microscopic many-body theory of radio-frequency dimer-projection spectroscopy remains largely unexplored. Existing work has focused on the short-pulse regime, where only the integrated dimer weight is measured. Much less is known about the frequency-resolved spectrum in the Fermi's golden rule regime. In particular, several fundamental questions remain open. How does many-body pairing modify the dimer spectrum? Does the spectrum contain information beyond the contact? Can dimer projection distinguish between short-range pair correlations and long-range superfluid coherence \cite{Yang1962}? Answering these questions is essential for understanding the full capability of this emerging spectroscopic technique.

In this work, we develop a microscopic theory of radio-frequency dimer-projection spectroscopy for a unitary Fermi gas. Exploiting the fact that the final-state dimer is much smaller than the interparticle spacing, we introduce a point-dimer approximation (PDA) that expresses the RF response directly in terms of pair correlations of the initial many-body state. We derive the frequency-dependent spectrum within the BCS mean-field theory, and identify two qualitatively distinct contributions. Besides a broad continuum, due to creation and annihilation of quasiparticle pairs and scattering of single quasiparticles, whose structure depends on the BCS pairing gap, the spectrum exhibits a sharp \emph{pair-condensation peak}, arising from the coherent projection of condensed Cooper pairs onto the final-state dimer. This coherent transfer process is analogous to the dc Josephson effect, in which Cooper pairs tunnel coherently between two superconductors~\cite{Josephson1962}, and therefore provides a direct spectroscopic manifestation of macroscopic phase coherence.

It is worth pointing out that, although conventional RF spectroscopy can also contain information about pair condensation, such information is strongly convolved with other effects in the one-body spectral function, and therefore does not generally produce a distinct spectral feature. By contrast, in our scheme, the constraint imposed by the projected final-state dimer allows the spectroscopy to effectively probe the two-body spectral function [see Eq.~(\ref{eq:spectrum})]. In the unitary regime, pair condensation gives rise to a prominent and narrow peak in this spectral function, providing a direct and distinctive spectroscopic signature of condensation. 

Therefore, our theory reveals that radio-frequency dimer-projection spectroscopy unifies several fundamental properties of strongly interacting Fermi gases within a single experimental observable. While the integrated spectral weight reproduces the recently discovered contact relation, the continuum involving quasiparticles provides information on the pairing gap, and the pair-condensation peak provides a direct probe of pair condensation and macroscopic coherence. The observable therefore simultaneously accesses short-range correlations, fermionic pairing, and long-range coherence, establishing dimer-projection spectroscopy as a powerful new tool for investigating equilibrium and nonequilibrium many-body physics.

\emph{\textcolor{blue}{Point-dimer approximation.---}}We consider a spin-balanced two-component Fermi gas of
$^{40}$K atoms in the hyperfine states
$|1\rangle$ and $|2\rangle$,
tuned near the broad $s$-wave Feshbach resonance at
$B_0\simeq202.14~{\rm G}$.
A radio-frequency (RF) field drives transitions from
state $|2\rangle$ to an initially unoccupied state
$|3\rangle$ through the operator
\begin{equation}
M=\int d{\bf r}\,
\psi_3^\dagger({\bf r})\psi_2({\bf r}),
\end{equation}
where $\psi_\sigma({\bf r})$ annihilates an atom in
hyperfine state $|\sigma\rangle$.
Throughout this work we consider the weak-driving and sufficiently long duration 
(Fermi's golden rule) regime
\cite{Baym2007,Punk2007,Zhang2008}, where the transition rate is given by

\begin{equation}
\Gamma(\omega)
=
2\pi
\sum_{i,f}
\rho_i
|\langle f|M|i\rangle|^2
\delta(\omega+E_i-E_f),
\label{eq:FGR}
\end{equation}
with $|i\rangle$ and $|f\rangle$ denoting the initial
and final many-body eigenstates and $\rho_i$ the
thermal distribution of the initial state.
The detuning $\omega$ is measured relative to the bare
$2\rightarrow3$ hyperfine transition frequency, and
we set $\hbar=1$ throughout. Note that  we have absorbed the RF coupling amplitude into the definition of the transition rate and report $\Gamma(\omega)$ in units of its squared magnitude.

The key observation underlying our theory is a clear
separation of energy and length scales.
At the magnetic field employed in the recent
dimer-projection experiment
\cite{Xie2026},
the final state $1$--$3$ channel supports a
bound dimer with binding energy
$|E_d|\simeq 2\pi\times4~{\rm MHz}$,
whereas the characteristic many-body energy scale is
the Fermi energy
$E_F\simeq 2\pi\times10~{\rm kHz}$.
Consequently,
\begin{equation}
|E_d|\gg E_F,
\qquad
\kappa=\sqrt{m|E_d|}\gg k_F,
\end{equation}
so that the dimer size
$\kappa^{-1}$ is much smaller than both the mean
interparticle spacing $k_F^{-1}$ ($E_F=k_F^2/2m$) and the Cooper-pair
size.
The newly formed dimer therefore behaves as an
essentially point-like composite object on the length
scales relevant to the many-body gas. Away from any additional few-body resonances involving the atom–dimer channel, the low-energy
atom--dimer scattering length is expected to be of the
order of the dimer size, $a_{ad}\sim\kappa^{-1}$
\cite{Petrov2004,BraatenHammer2006}. Interactions between the
dimer and the surrounding atoms then produce corrections
controlled parametrically by $k_Fa_{ad}$. Motivated by
the hierarchy $k_F/\kappa\ll1$, we introduce the
point-dimer approximation, in which these residual
atom--dimer interactions are neglected to leading order.

Accordingly, the final many-body state factorizes into
a free dimer carrying center-of-mass momentum ${\bf P}$
and the remaining many-body state,
\begin{equation}
|f\rangle
=
\int
d{\bf R}\,d{\bf r}\,
\frac{e^{i{\bf P}\cdot{\bf R}}}{\sqrt V}
\phi_d({\bf r})
\psi_1^\dagger
({\bf R}+{\bf r}/2)
\psi_3^\dagger
({\bf R}-{\bf r}/2)
|f'\rangle,
\label{eq:finalstate}
\end{equation}
where $\phi_d({\bf r})$ denotes the dimer relative
wave-function whose form is $\sqrt{\kappa/2\pi} e^{-\kappa r}/r$
outside the interaction range, $V$ is the system volume, 
and $|f'\rangle$ is the state of the remaining
$|1\rangle$ and $|2\rangle$ atoms.
Substituting Eq.~(\ref{eq:finalstate}) into
Eq.~(\ref{eq:FGR}) gives
\begin{align}
\Gamma(\omega)
=&\,
2\pi
\sum_{i,f',{\bf P}}
\rho_i
\Bigg|
\frac1{\sqrt V}
\sum_{\bf q}
\tilde\phi_d^*({\bf q})
\langle f'|
a_{1,{\bf P}/2+{\bf q}}
a_{2,{\bf P}/2-{\bf q}}
|i\rangle
\Bigg|^2
\nonumber\\
&
\times
\delta
\!\left(
\omega+E_i-E_{f'}
+|E_d|
-\frac{P^2}{4m}
\right),
\label{eq:spectrum}
\end{align}
with the Fourier transforms $a_{\sigma,\mathbf k}= \int d\mathbf{r}\,\psi_\sigma(\mathbf{r})e^{-i\mathbf k\cdot\mathbf r}/\sqrt V$ and $\tilde{\phi}_d(\mathbf{q})= \int d\mathbf{r}\,\phi_d(\mathbf{r})e^{-i\mathbf q\cdot\mathbf r}$.

Equation~(\ref{eq:spectrum}) is the central result of
the point-dimer approximation.
Unlike conventional RF spectroscopy, which probes
single-particle excitations, 
dimer-projection spectroscopy probes matrix elements of the pair-annihilation operator weighted by the final-state molecular wave-function. Therefore the RF response becomes a direct probe of
two-particle correlations rather than single-particle
spectral functions, providing the foundation for the
analysis presented below.

Given that Eq.~(\ref{eq:spectrum}) is independent of any specific many-body approximation, its evaluation can be combined with a variety of theoretical descriptions of the initial state. In the following, we employ the BCS mean-field theory as the simplest realization to illustrate the physical consequences of the point-dimer approximation and the microscopic origin of the features of the dimer-projection spectrum.

\emph{\textcolor{blue}{Mean-field.---}}
To elucidate how many-body pairing manifests itself in the dimer-projection spectrum, we describe the interacting 1–2 Fermi gas within the BCS mean-field theory. Although quantitatively approximate at unitarity, the
BCS theory is known to capture essential ingredients of
fermionic superfluidity~\cite{BCS1957,Leggett1980,Giorgini2008,Haussmann2009}.
More importantly, it provides a transparent microscopic
interpretation of the dimer-projection spectrum by
decomposing the pair annihilation operator [see Eq.~(\ref{eq:pairop})] into several
elementary transfer processes. See the Supplementary Material for derivation details \cite{SM}.
We argue that qualitative features identified below are expected to persist beyond the BCS theory.




The BCS mean-field Hamiltonian for the interacting 1--2 Fermi gas can be diagonalized in terms of
the quasiparticles whose annihilation operators $\gamma_{\sigma,\mathbf{k}}$ are given by the Bogoliubov transformation \cite{Giorgini2008}
\begin{equation}
\begin{pmatrix}
a_{1,\mathbf{k}}
\\
a_{2,-\mathbf{k}}^\dagger
\end{pmatrix}
=
\begin{pmatrix}
-\cos\theta_{\mathbf{k}}
&
\sin\theta_{\mathbf{k}}
\\
\sin\theta_{\mathbf{k}}
&
\cos\theta_{\mathbf{k}}
\end{pmatrix}
\begin{pmatrix}
\gamma_{1,\mathbf{k}}
\\
\gamma_{2,-\mathbf{k}}^\dagger
\end{pmatrix},
\label{eq:Bogo}
\end{equation}
where
\[
\cos\theta_{\mathbf{k}}
=
\sqrt{\frac{E_{\mathbf{k}}+\xi_{\mathbf{k}}}
{2E_{\mathbf{k}}}},
\qquad
\sin\theta_{\mathbf{k}}
=
\sqrt{\frac{\Delta^2}
{2E_{\mathbf{k}}
(E_{\mathbf{k}}+\xi_{\mathbf{k}})}},
\]
and $E_{\mathbf{k}}
=
\sqrt{\xi_{\mathbf{k}}^2+\Delta^2}$, $\xi_{\mathbf{k}}=k^2/2m-\mu$, $\Delta$ is the BCS pairing gap, and $\mu$ is the chemical potential.

The key quantity entering the point-dimer approximation
is the pair annihilation operator
\[
a_{1,\mathbf{P}/2+\mathbf q}
a_{2,\mathbf{P}/2-\mathbf q},
\]
which removes a pair of atoms from the initial
many-body state. Expressing it in terms of Bogoliubov quasiparticles
yields
\begin{align}
&a_{1,\mathbf{P}/2+\mathbf{q}} a_{2,\mathbf{P}/2-\mathbf{q}}
\nonumber\\
=&
[-\cos\theta_+ \gamma_{1,\mathbf{P}/2+\mathbf{q}} + \sin\theta_+ \gamma_{2,-(\mathbf{P}/2+\mathbf{q})}^\dagger]\nonumber \\
&\times [\sin\theta_- \gamma_{1,-(\mathbf{P}/2-\mathbf{q})}^\dagger + \cos\theta_- \gamma_{2,\mathbf{P}/2-\mathbf{q}}],
\label{eq:pairop}
\end{align}
where
$\theta_\pm
=
\theta_{|\mathbf P/2\pm\mathbf q|}$.
Equation~(\ref{eq:pairop}) provides a physical interpretation of radio-frequency dimer-projection
spectroscopy.
There are four classes of terms corresponding to four elementary transfer
processes:

(i) coherent transfer of condensed Cooper pairs,

(ii) creation of two Bogoliubov quasiparticles,

(iii) annihilation of two excited
Bogoliubov quasiparticles,

(iv) scattering of a single excited Bogoliubov quasiparticle.

\begin{figure}[t]
\includegraphics[width=\columnwidth]{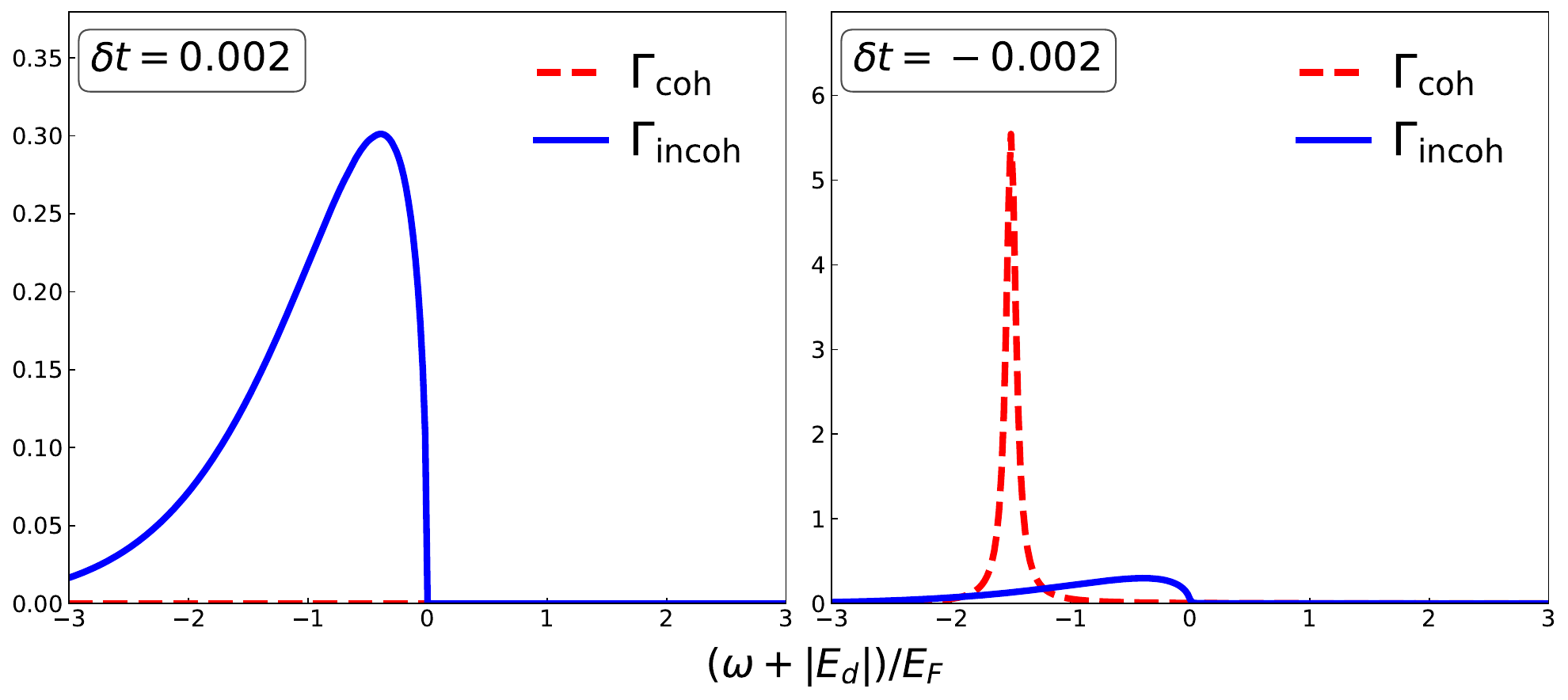}
\caption{\label{spectrum} 
The RF dimer-projection spectrum components $\Gamma_{\rm coh}(\omega)$ and $\Gamma_{\rm incoh}(\omega)$ at unitarity for different $\delta t=(T-T_c)/T_c$. Here $T_c$ is the BCS transition temperature. We assume that the pair-condensation peak is broadened as a Lorentzian, Eq.~(\ref{loren}), with $1/\tau_m=E_F/10$. Numerical evaluations of $\Gamma_{\rm coh}(\omega)$ and $\Gamma_{\rm incoh}(\omega)$ are within the BCS mean-field theory and assume the dimer wave-function $\tilde\phi_d(\mathbf q)=\sqrt{8\pi\kappa}/(q^2+\kappa^2)$ with $\kappa/k_F=10\sqrt2$ corresponding to the experimental value. The unit of the two components in the plot is $3N_{\rm tot}|\tilde\phi_d(0)|^2 mk_F/4\pi$ with $N_{\rm tot}$ the total number of atoms.}
\end{figure}

Since these processes lead to mutually orthogonal final
states, the RF spectrum naturally decomposes into
independent contributions as
\begin{align}
\Gamma(\omega)
=&
\Gamma_{\rm coh}(\omega)
+
\Gamma_{\rm incoh}(\omega),\label{eq:GammaDecomp}\\
\Gamma_{\rm incoh}(\omega)=&
\Gamma_{\rm create}(\omega)
+
\Gamma_{\rm annih}(\omega)
+
\Gamma_{\rm scatter}(\omega).
\label{eq:GammaIncoh}
\end{align}
Figure \ref{spectrum} shows the BCS mean-field results for $\Gamma_{\rm coh}(\omega)$ and $\Gamma_{\rm incoh}(\omega)$ at unitarity as the temperature $T$ lowers through the superfluid transition.

The coherent contribution originates from the direct
projection of condensed Cooper pairs onto the final-state
dimer,
\begin{align}
\Gamma_{\rm coh}(\omega)
=
& \frac{2\pi}{V} \left| \sum_{\mathbf{q}} \tilde{\phi}_d(\mathbf{q}) \cos\theta_{q} \sin\theta_{q} [1 - 2f(E_{q})] \right|^2
\nonumber\\
&\times\delta(\omega + |E_d|+2\mu).\label{eq:Gammacoh}
\end{align}
Here the Fermi distribution is $f(x) = 1/(e^{x/T} + 1)$. 
Unlike all remaining contributions, this process does
not involve quasiparticles. It therefore gives rise to a sharp \emph{pair-condensation peak}, which constitutes the spectroscopic signature of macroscopic phase coherence in the superfluid.
The factor $1-2f(E_q)$ accounts for the depletion of the coherent pair amplitude by thermally excited quasiparticles. 
The physical meaning of this peak frequency $-(|E_d| + 2\mu)$ follows from the coherent-transfer process. Extracting a pair of atoms from the Fermi gas at fixed entropy lowers the energy of the gas by $2\mu$, while converting the transferred pair into a bound dimer gains an additional energy $|E_d|$. As a result, the final state is lower in energy than the initial state by $|E_d| + 2\mu$, so the peak is red-shifted from the bare atomic transition frequency by exactly this amount.

It is worth emphasizing that the existence of the pair-condensation peak is unlikely to be a peculiarity of the BCS approximation. Rather, it follows
directly from the coherent transfer of atom pairs into
the final-state dimer and is therefore a generic
consequence of off-diagonal long-range order in the
initial fermionic superfluid state~\cite{Yang1962}. Equation~(\ref{eq:spectrum}), which constitutes
the central result of the point-dimer approximation,
expresses the RF response solely in terms of the pair
correlation function and is independent of the
many-body approximation employed to evaluate it.
Whenever the two-body density matrix of the initial state possesses a macroscopic eigenvalue, its condensate contribution is anticipated to produce a coherent dimer-projection peak whose weight depends on the overlap between the corresponding pair eigenfunction and the final-state molecular wave-function.
Different many-body theories may modify its
spectral weight
through quantitative
changes 
in the pair correlations, but not its physical
origin.

The coherent transfer process bears a close analogy to
the dc Josephson effect \cite{Josephson1962}. In a Josephson junction,
Cooper pairs tunnel coherently between two
superconductors without generating quasiparticle
excitations. Likewise, the RF field coherently converts
a condensed Cooper pair into a bound molecular state,
again bypassing quasiparticle excitations. 
Unlike a conventional Josephson junction, the present process involves only one condensate, and the coherent pair transfer is driven by the RF field rather than a phase difference across a junction. The analogy therefore lies in the coherent transfer of condensed pairs without quasiparticle excitations.


In experiments, the narrow peak is broadened by several
mechanisms. In particular, the resulting 1--3 dimers
undergo inelastic collisions with atoms in state $|2\rangle$,
leading to a finite lifetime $\tau_m\sim10/E_F$ \cite{Xie2026}.
Accordingly, we replace
$2\pi\delta(\omega+|E_d|+2\mu)$ in
$\Gamma_{\rm coh}(\omega)$ by the Lorentzian
\begin{align}
\frac{1/\tau_m}
{(\omega+|E_d|+2\mu)^2+(1/2\tau_m)^2}.\label{loren}
\end{align}
Figure~\ref{spectrum} 
shows that the broadened peak remains clearly resolvable and rapidly dominates the maximum spectral response as the temperature is lowered below the superfluid transition temperature $T_c$.

The remaining three transfer processes all involve quasiparticles which carry away phase information. The corresponding terms are given by 
\begin{align}
\Gamma_{\rm create}(\omega)=
&\frac{2\pi}{V} \sum_{\mathbf{P},\mathbf{q}} |\tilde{\phi}_d(\mathbf{q})|^2 \sin^2\theta_+ \sin^2\theta_- f(-E_+) f(-E_-) 
\nonumber\\
&\times\delta(\omega - E_+ - E_- + |E_d| - P^2/4m+2\mu),
\label{eq:Gammacreate}
\\ 
\Gamma_{\rm annih}(\omega) =
& \frac{2\pi}{V} \sum_{\mathbf{P},\mathbf{q}} |\tilde{\phi}_d(\mathbf{q})|^2 \cos^2\theta_+ \cos^2\theta_- f(E_+) f(E_-) 
\nonumber\\
&\times \delta(\omega + E_+ + E_- + |E_d| - P^2/4m+2\mu), 
\label{eq:Gammaannih}
\\
\Gamma_{\rm scatter}(\omega) =
& \frac{4\pi}{V} \sum_{\mathbf{P},\mathbf{q}} |\tilde{\phi}_d(\mathbf{q})|^2 \cos^2\theta_+ \sin^2\theta_- f(-E_-) f(E_+)
\nonumber\\
&\times\delta(\omega + E_+ - E_- + |E_d| - P^2/4m+2\mu).
\label{eq:Gammascatter}
\end{align}
In these sums, the subscripts $\pm$ denote $\mathbf P/2\pm\mathbf q$ respectively. The combined contributions of the three terms give rise to a spread distribution $\Gamma_{\rm incoh}(\omega)$ of width $\sim E_F$ as shown in Fig.~\ref{spectrum}. 
Thermally excited quasiparticles are required for both $\Gamma_{\rm annih}(\omega)$ and $\Gamma_{\rm scatter}(\omega)$ to be nonzero.
In particular, at $T=0$, of these three terms, only $\Gamma_{\rm create}(\omega)$ still contributes, and it has a threshold at $\omega=-2\mu-|E_d|+2\Delta$; 
thus the continuum threshold is separated from the coherent peak by  $2\Delta$.

Numerical evaluations of $\Gamma_{\rm coh}(\omega)$ and $\Gamma_{\rm incoh}(\omega)$ within the BCS mean-field theory, whose results are plotted in Fig.~\ref{spectrum}, assume the dimer wave-function $\tilde\phi_d(\mathbf q)=\sqrt{8\pi\kappa}/(q^2+\kappa^2)$ with $\kappa/k_F=10\sqrt2$ corresponding to the experimental value. We have verified that the qualitative spectral structure is insensitive to the precise form of the molecular wave-function provided $\kappa\gg k_F$.

\emph{\textcolor{blue}{Sum rule.---}}Our theory naturally reproduces the recent experimental relation between integrated dimer weight and Tan's contact $C$~\cite{Xie2026}. This relation can be readily shown from Eq.~(\ref{eq:spectrum}), which yields the dimer weight as
\begin{align}
&\int \frac{d\omega}{2\pi} \Gamma(\omega) \nonumber\\
=& \int d\mathbf{R} \int d\mathbf{r} \int d\mathbf{r'}\, \phi_d(\mathbf{r}) \phi_d^*(\mathbf{r'})\rho(\mathbf R|\mathbf r, \mathbf r'),\label{int}
\end{align}
with the two-body density matrix 
\begin{align}
\rho(\mathbf R|\mathbf r, \mathbf r')\equiv \langle \psi_2^\dagger(\mathbf R-\frac{\mathbf{r}}2) \psi_1^\dagger(\mathbf R+\frac{\mathbf{r}}2) \psi_1(\mathbf R+\frac{\mathbf{r'}}2) \psi_2(\mathbf R-\frac{\mathbf{r'}}2) \rangle,
\end{align}
and $\langle\dots\rangle$ standing for $\sum_i\rho_i\langle i|\dots|i\rangle$.
Because the two-body density matrix is hermitian, i.e., $\rho(\mathbf R|\mathbf r, \mathbf r')=\rho^*(\mathbf R|\mathbf r', \mathbf r)$, there exists a complete set of normalized and orthogonal eigenstates $\{\chi_j(\mathbf R,\mathbf r)\}$ and eigenvalues $\{\lambda_j(\mathbf R)\}$ satisfying 
$\int d\mathbf r' \rho(\mathbf R|\mathbf r, \mathbf r') \chi_j(\mathbf R,\mathbf r')=\lambda_j(\mathbf R) \chi_j(\mathbf R,\mathbf r)$.  
Correspondingly, one can express the two-body density matrix as
$\rho(\mathbf R|\mathbf r, \mathbf r')=\sum_j \lambda_j(\mathbf R) \chi_j(\mathbf R,\mathbf r)\chi_j^*(\mathbf R,\mathbf r')$.
Due to the diluteness of atomic gases, when two interacting atoms approach each other at a distance $r$ that is much smaller than the mean interparticle spacing $d$, the eigenstates have the asymptotic form
$\chi_j(\mathbf R,\mathbf r)\sim A_j(\mathbf R)\tilde\chi(\mathbf r)$. Here $\tilde\chi(\mathbf r)$ is the zero energy two-body relative wave-function, and has an asymptotic form $\sim 1/r - 1/a_{12}$ in the regime $r_0\ll r\ll d$, with $r_0$ the interaction range and $a_{12}$ the 1--2 atomic scattering length~\cite{Tan2008a, Tan2008b, Tan2008c, Zhang2009}.
Thus, given that the size of the dimer $\phi_d(\mathbf r)$ in Eq.~(\ref{int}) is much smaller than $d$, the weight becomes
\begin{align}
\int \frac{d\omega}{2\pi} \Gamma(\omega)=C\left|\int d\mathbf r \tilde\chi(\mathbf r)\phi_d^*(\mathbf r)\right|^2,
\end{align}
with the extensive contact \cite{Zhang2009}
\begin{align}
C=\int d\mathbf R \sum_j \lambda_j(\mathbf R)|A_j(\mathbf R)|^2;
\end{align}
the overlap integral between $\tilde\chi(\mathbf r)$ and $\phi_d(\mathbf r)$ encodes the dependence on short range physics. 

\emph{\textcolor{blue}{Discussion.---}}We have developed a microscopic theory of radio-frequency
dimer-projection spectroscopy for strongly interacting
Fermi gases. Exploiting the large separation between the
molecular and many-body length scales, we introduced the
point-dimer approximation (PDA), which expresses the
radio-frequency response directly in terms of pair
correlations of the initial many-body state. The PDA
provides a simple and physically transparent framework
for describing final-state interactions in
dimer-projection spectroscopy.

Within the BCS mean-field theory, we showed that the
radio-frequency dimer-projection spectrum naturally decomposes into four
elementary transfer processes. Besides a broad
quasiparticle continuum associated with pair creation,
pair annihilation, and quasiparticle scattering, the
spectrum exhibits a sharp pair-condensation peak arising
from the coherent projection of condensed Cooper pairs
onto the final-state dimer. The continuum possesses a
threshold determined by the superfluid gap in the low-temperature limit, whereas the
pair-condensation peak directly reflects the macroscopic
phase coherence of the superfluid. These two distinct
spectral structures therefore provide complementary
signatures of fermionic pairing and condensation.

Treatments beyond the BCS theory, such as T-matrix or pair-fluctuation theories incorporating collective modes \cite{LiuHu2005,KliminTempereKurkjian2019}, could modify the spectrum through finite quasiparticle lifetimes \cite{VanLoonTempereKurkjian2020}, pseudogap effects \cite{TsuchiyaWatanabeOhashi2009,GaeblerEtAl2010}, and additional collective excitations \cite{CombescotKaganStringari2006,KurkjianKliminTempereCastin2019}, thereby smearing the sharp BCS threshold and redistributing spectral weight \cite{Haussmann2009,PeraliPieriStrinati2008}. However, at the same time, we expect the existence of a coherent dimer-projection peak to be qualitatively robust, since it is associated with off-diagonal long-range order rather than with any special artifact of mean-field theory~\cite{Yang1962}. 

A recent proposal suggests employing a Rydberg atom to detect pairing and condensation in unitary Fermi gases \cite{RamosRodriguez2026}. The Rydberg atom creates a local potential that supports one-body and two-body bound states, and transferring atoms into the two-body bound states is predicted to provide a spectroscopic signature of pair condensation. This approach is closely related to ours in that both exploit a final two-body bound state to achieve the sensitivity to pair condensation. However, the physical implementations are different: the Rydberg-based scheme relies on an additional localized Rydberg potential, whereas our scheme directly employs a projected final-state dimer within the RF spectroscopy. 

\begin{acknowledgments}
This work was supported by the Quantum Science and Technology-National Science and Technology Major Project, China (Grant
No. 2025ZD0300400) and the National Natural Science Foundation of China (Grant No. 12474270).
\end{acknowledgments}

\bibliography{ref}

\providecommand{\noopsort}[1]{}\providecommand{\singleletter}[1]{#1}%
\begin{thebibliography}{58}
\expandafter\ifx\csname natexlab\endcsname\relax\def\natexlab#1{#1}\fi
\expandafter\ifx\csname bibnamefont\endcsname\relax
  \def\bibnamefont#1{#1}\fi
\expandafter\ifx\csname bibfnamefont\endcsname\relax
  \def\bibfnamefont#1{#1}\fi
\expandafter\ifx\csname citenamefont\endcsname\relax
  \def\citenamefont#1{#1}\fi
\expandafter\ifx\csname url\endcsname\relax
  \def\url#1{\texttt{#1}}\fi
\expandafter\ifx\csname urlprefix\endcsname\relax\def\urlprefix{URL }\fi
\providecommand{\bibinfo}[2]{#2}
\providecommand{\eprint}[2][]{\url{#2}}

\bibitem[{\citenamefont{Gupta et~al.}(2003)\citenamefont{Gupta, Hadzibabic,
  Zwierlein, Stan, Dieckmann, Schunck, van Kempen, Verhaar, and
  Ketterle}}]{Gupta2003}
\bibinfo{author}{\bibfnamefont{S.}~\bibnamefont{Gupta}},
  \bibinfo{author}{\bibfnamefont{Z.}~\bibnamefont{Hadzibabic}},
  \bibinfo{author}{\bibfnamefont{M.~W.} \bibnamefont{Zwierlein}},
  \bibinfo{author}{\bibfnamefont{C.~A.} \bibnamefont{Stan}},
  \bibinfo{author}{\bibfnamefont{K.}~\bibnamefont{Dieckmann}},
  \bibinfo{author}{\bibfnamefont{C.~H.} \bibnamefont{Schunck}},
  \bibinfo{author}{\bibfnamefont{E.~G.~M.} \bibnamefont{van Kempen}},
  \bibinfo{author}{\bibfnamefont{B.~J.} \bibnamefont{Verhaar}},
  \bibnamefont{and} \bibinfo{author}{\bibfnamefont{W.}~\bibnamefont{Ketterle}},
  \bibinfo{journal}{Science} \textbf{\bibinfo{volume}{300}},
  \bibinfo{pages}{1723} (\bibinfo{year}{2003}),
  \urlprefix\url{https://www.science.org/doi/abs/10.1126/science.1085335}.

\bibitem[{\citenamefont{Chin et~al.}(2004)\citenamefont{Chin, Bartenstein,
  Altmeyer, Riedl, Jochim, Hecker~Denschlag, and Grimm}}]{Chin2004}
\bibinfo{author}{\bibfnamefont{C.}~\bibnamefont{Chin}},
  \bibinfo{author}{\bibfnamefont{M.}~\bibnamefont{Bartenstein}},
  \bibinfo{author}{\bibfnamefont{A.}~\bibnamefont{Altmeyer}},
  \bibinfo{author}{\bibfnamefont{S.}~\bibnamefont{Riedl}},
  \bibinfo{author}{\bibfnamefont{S.}~\bibnamefont{Jochim}},
  \bibinfo{author}{\bibfnamefont{J.}~\bibnamefont{Hecker~Denschlag}},
  \bibnamefont{and} \bibinfo{author}{\bibfnamefont{R.}~\bibnamefont{Grimm}},
  \bibinfo{journal}{Science} \textbf{\bibinfo{volume}{305}},
  \bibinfo{pages}{1128} (\bibinfo{year}{2004}),
  \urlprefix\url{https://www.science.org/doi/abs/10.1126/science.1100818}.

\bibitem[{\citenamefont{Schunck et~al.}(2007)\citenamefont{Schunck, Shin,
  Schirotzek, and Ketterle}}]{Schunck2007}
\bibinfo{author}{\bibfnamefont{C.~H.} \bibnamefont{Schunck}},
  \bibinfo{author}{\bibfnamefont{Y.}~\bibnamefont{Shin}},
  \bibinfo{author}{\bibfnamefont{A.}~\bibnamefont{Schirotzek}},
  \bibnamefont{and} \bibinfo{author}{\bibfnamefont{W.}~\bibnamefont{Ketterle}},
  \bibinfo{journal}{Science} \textbf{\bibinfo{volume}{316}},
  \bibinfo{pages}{867} (\bibinfo{year}{2007}),
  \urlprefix\url{https://www.science.org/doi/abs/10.1126/science.1140749}.

\bibitem[{\citenamefont{Schirotzek et~al.}(2008)\citenamefont{Schirotzek, Shin,
  Schunck, and Ketterle}}]{Schirotzek2008}
\bibinfo{author}{\bibfnamefont{A.}~\bibnamefont{Schirotzek}},
  \bibinfo{author}{\bibfnamefont{Y.-i.} \bibnamefont{Shin}},
  \bibinfo{author}{\bibfnamefont{C.~H.} \bibnamefont{Schunck}},
  \bibnamefont{and} \bibinfo{author}{\bibfnamefont{W.}~\bibnamefont{Ketterle}},
  \bibinfo{journal}{Phys. Rev. Lett.} \textbf{\bibinfo{volume}{101}},
  \bibinfo{pages}{140403} (\bibinfo{year}{2008}),
  \urlprefix\url{https://link.aps.org/doi/10.1103/PhysRevLett.101.140403}.

\bibitem[{\citenamefont{Mukherjee et~al.}(2019)\citenamefont{Mukherjee, Patel,
  Yan, Fletcher, Struck, and Zwierlein}}]{Mukherjee2019}
\bibinfo{author}{\bibfnamefont{B.}~\bibnamefont{Mukherjee}},
  \bibinfo{author}{\bibfnamefont{P.~B.} \bibnamefont{Patel}},
  \bibinfo{author}{\bibfnamefont{Z.}~\bibnamefont{Yan}},
  \bibinfo{author}{\bibfnamefont{R.~J.} \bibnamefont{Fletcher}},
  \bibinfo{author}{\bibfnamefont{J.}~\bibnamefont{Struck}}, \bibnamefont{and}
  \bibinfo{author}{\bibfnamefont{M.~W.} \bibnamefont{Zwierlein}},
  \bibinfo{journal}{Phys. Rev. Lett.} \textbf{\bibinfo{volume}{122}},
  \bibinfo{pages}{203402} (\bibinfo{year}{2019}),
  \urlprefix\url{https://link.aps.org/doi/10.1103/PhysRevLett.122.203402}.

\bibitem[{\citenamefont{Stewart et~al.}(2008)\citenamefont{Stewart, Gaebler,
  and Jin}}]{Stewart2008}
\bibinfo{author}{\bibfnamefont{J.~T.} \bibnamefont{Stewart}},
  \bibinfo{author}{\bibfnamefont{J.~P.} \bibnamefont{Gaebler}},
  \bibnamefont{and} \bibinfo{author}{\bibfnamefont{D.~S.} \bibnamefont{Jin}},
  \bibinfo{journal}{Nature} \textbf{\bibinfo{volume}{454}},
  \bibinfo{pages}{744} (\bibinfo{year}{2008}), ISSN \bibinfo{issn}{1476-4687},
  \urlprefix\url{https://doi.org/10.1038/nature07172}.

\bibitem[{\citenamefont{Vale and Zwierlein}(2021)}]{Vale2021}
\bibinfo{author}{\bibfnamefont{C.~J.} \bibnamefont{Vale}} \bibnamefont{and}
  \bibinfo{author}{\bibfnamefont{M.}~\bibnamefont{Zwierlein}},
  \bibinfo{journal}{Nature Physics} \textbf{\bibinfo{volume}{17}},
  \bibinfo{pages}{1305} (\bibinfo{year}{2021}), ISSN \bibinfo{issn}{1745-2481},
  \urlprefix\url{https://doi.org/10.1038/s41567-021-01434-6}.

\bibitem[{\citenamefont{Tan}(2008{\natexlab{a}})}]{Tan2008a}
\bibinfo{author}{\bibfnamefont{S.}~\bibnamefont{Tan}}, \bibinfo{journal}{Annals
  of Physics} \textbf{\bibinfo{volume}{323}}, \bibinfo{pages}{2952}
  (\bibinfo{year}{2008}{\natexlab{a}}), ISSN \bibinfo{issn}{0003-4916},
  \urlprefix\url{https://www.sciencedirect.com/science/article/pii/S0003491608000456}.

\bibitem[{\citenamefont{Tan}(2008{\natexlab{b}})}]{Tan2008b}
\bibinfo{author}{\bibfnamefont{S.}~\bibnamefont{Tan}}, \bibinfo{journal}{Annals
  of Physics} \textbf{\bibinfo{volume}{323}}, \bibinfo{pages}{2971}
  (\bibinfo{year}{2008}{\natexlab{b}}), ISSN \bibinfo{issn}{0003-4916},
  \urlprefix\url{https://www.sciencedirect.com/science/article/pii/S0003491608000432}.

\bibitem[{\citenamefont{Tan}(2008{\natexlab{c}})}]{Tan2008c}
\bibinfo{author}{\bibfnamefont{S.}~\bibnamefont{Tan}}, \bibinfo{journal}{Annals
  of Physics} \textbf{\bibinfo{volume}{323}}, \bibinfo{pages}{2987}
  (\bibinfo{year}{2008}{\natexlab{c}}), ISSN \bibinfo{issn}{0003-4916},
  \urlprefix\url{https://www.sciencedirect.com/science/article/pii/S0003491608000420}.

\bibitem[{\citenamefont{Braaten et~al.}(2010)\citenamefont{Braaten, Kang, and
  Platter}}]{Braaten2010}
\bibinfo{author}{\bibfnamefont{E.}~\bibnamefont{Braaten}},
  \bibinfo{author}{\bibfnamefont{D.}~\bibnamefont{Kang}}, \bibnamefont{and}
  \bibinfo{author}{\bibfnamefont{L.}~\bibnamefont{Platter}},
  \bibinfo{journal}{Phys. Rev. Lett.} \textbf{\bibinfo{volume}{104}},
  \bibinfo{pages}{223004} (\bibinfo{year}{2010}),
  \urlprefix\url{https://link.aps.org/doi/10.1103/PhysRevLett.104.223004}.

\bibitem[{\citenamefont{Schneider and Randeria}(2010)}]{Schneider2010}
\bibinfo{author}{\bibfnamefont{W.}~\bibnamefont{Schneider}} \bibnamefont{and}
  \bibinfo{author}{\bibfnamefont{M.}~\bibnamefont{Randeria}},
  \bibinfo{journal}{Phys. Rev. A} \textbf{\bibinfo{volume}{81}},
  \bibinfo{pages}{021601(R)} (\bibinfo{year}{2010}),
  \urlprefix\url{https://link.aps.org/doi/10.1103/PhysRevA.81.021601}.

\bibitem[{\citenamefont{Stewart et~al.}(2010)\citenamefont{Stewart, Gaebler,
  Drake, and Jin}}]{Stewart2010}
\bibinfo{author}{\bibfnamefont{J.~T.} \bibnamefont{Stewart}},
  \bibinfo{author}{\bibfnamefont{J.~P.} \bibnamefont{Gaebler}},
  \bibinfo{author}{\bibfnamefont{T.~E.} \bibnamefont{Drake}}, \bibnamefont{and}
  \bibinfo{author}{\bibfnamefont{D.~S.} \bibnamefont{Jin}},
  \bibinfo{journal}{Phys. Rev. Lett.} \textbf{\bibinfo{volume}{104}},
  \bibinfo{pages}{235301} (\bibinfo{year}{2010}),
  \urlprefix\url{https://link.aps.org/doi/10.1103/PhysRevLett.104.235301}.

\bibitem[{\citenamefont{Sagi et~al.}(2012)\citenamefont{Sagi, Drake, Paudel,
  and Jin}}]{Sagi2012}
\bibinfo{author}{\bibfnamefont{Y.}~\bibnamefont{Sagi}},
  \bibinfo{author}{\bibfnamefont{T.~E.} \bibnamefont{Drake}},
  \bibinfo{author}{\bibfnamefont{R.}~\bibnamefont{Paudel}}, \bibnamefont{and}
  \bibinfo{author}{\bibfnamefont{D.~S.} \bibnamefont{Jin}},
  \bibinfo{journal}{Phys. Rev. Lett.} \textbf{\bibinfo{volume}{109}},
  \bibinfo{pages}{220402} (\bibinfo{year}{2012}),
  \urlprefix\url{https://link.aps.org/doi/10.1103/PhysRevLett.109.220402}.

\bibitem[{\citenamefont{Shkedrov et~al.}(2018)\citenamefont{Shkedrov,
  Florshaim, Ness, Gandman, and Sagi}}]{Shkedrov2018}
\bibinfo{author}{\bibfnamefont{C.}~\bibnamefont{Shkedrov}},
  \bibinfo{author}{\bibfnamefont{Y.}~\bibnamefont{Florshaim}},
  \bibinfo{author}{\bibfnamefont{G.}~\bibnamefont{Ness}},
  \bibinfo{author}{\bibfnamefont{A.}~\bibnamefont{Gandman}}, \bibnamefont{and}
  \bibinfo{author}{\bibfnamefont{Y.}~\bibnamefont{Sagi}},
  \bibinfo{journal}{Phys. Rev. Lett.} \textbf{\bibinfo{volume}{121}},
  \bibinfo{pages}{093402} (\bibinfo{year}{2018}),
  \urlprefix\url{https://link.aps.org/doi/10.1103/PhysRevLett.121.093402}.

\bibitem[{\citenamefont{Baym et~al.}(2007)\citenamefont{Baym, Pethick, Yu, and
  Zwierlein}}]{Baym2007}
\bibinfo{author}{\bibfnamefont{G.}~\bibnamefont{Baym}},
  \bibinfo{author}{\bibfnamefont{C.~J.} \bibnamefont{Pethick}},
  \bibinfo{author}{\bibfnamefont{Z.}~\bibnamefont{Yu}}, \bibnamefont{and}
  \bibinfo{author}{\bibfnamefont{M.~W.} \bibnamefont{Zwierlein}},
  \bibinfo{journal}{Phys. Rev. Lett.} \textbf{\bibinfo{volume}{99}},
  \bibinfo{pages}{190407} (\bibinfo{year}{2007}),
  \urlprefix\url{https://link.aps.org/doi/10.1103/PhysRevLett.99.190407}.

\bibitem[{\citenamefont{Punk and Zwerger}(2007)}]{Punk2007}
\bibinfo{author}{\bibfnamefont{M.}~\bibnamefont{Punk}} \bibnamefont{and}
  \bibinfo{author}{\bibfnamefont{W.}~\bibnamefont{Zwerger}},
  \bibinfo{journal}{Phys. Rev. Lett.} \textbf{\bibinfo{volume}{99}},
  \bibinfo{pages}{170404} (\bibinfo{year}{2007}),
  \urlprefix\url{https://link.aps.org/doi/10.1103/PhysRevLett.99.170404}.

\bibitem[{\citenamefont{Zhang and Leggett}(2008)}]{Zhang2008}
\bibinfo{author}{\bibfnamefont{S.}~\bibnamefont{Zhang}} \bibnamefont{and}
  \bibinfo{author}{\bibfnamefont{A.~J.} \bibnamefont{Leggett}},
  \bibinfo{journal}{Phys. Rev. A} \textbf{\bibinfo{volume}{77}},
  \bibinfo{pages}{033614} (\bibinfo{year}{2008}),
  \urlprefix\url{https://link.aps.org/doi/10.1103/PhysRevA.77.033614}.

\bibitem[{\citenamefont{Zhang and Leggett}(2009)}]{Zhang2009}
\bibinfo{author}{\bibfnamefont{S.}~\bibnamefont{Zhang}} \bibnamefont{and}
  \bibinfo{author}{\bibfnamefont{A.~J.} \bibnamefont{Leggett}},
  \bibinfo{journal}{Phys. Rev. A} \textbf{\bibinfo{volume}{79}},
  \bibinfo{pages}{023601} (\bibinfo{year}{2009}),
  \urlprefix\url{https://link.aps.org/doi/10.1103/PhysRevA.79.023601}.

\bibitem[{\citenamefont{Langmack et~al.}(2012)\citenamefont{Langmack, Barth,
  Zwerger, and Braaten}}]{Langmack2012}
\bibinfo{author}{\bibfnamefont{C.}~\bibnamefont{Langmack}},
  \bibinfo{author}{\bibfnamefont{M.}~\bibnamefont{Barth}},
  \bibinfo{author}{\bibfnamefont{W.}~\bibnamefont{Zwerger}}, \bibnamefont{and}
  \bibinfo{author}{\bibfnamefont{E.}~\bibnamefont{Braaten}},
  \bibinfo{journal}{Phys. Rev. Lett.} \textbf{\bibinfo{volume}{108}},
  \bibinfo{pages}{060402} (\bibinfo{year}{2012}),
  \urlprefix\url{https://link.aps.org/doi/10.1103/PhysRevLett.108.060402}.

\bibitem[{\citenamefont{Fujii and Nishida}(2018)}]{Fujii18}
\bibinfo{author}{\bibfnamefont{K.}~\bibnamefont{Fujii}} \bibnamefont{and}
  \bibinfo{author}{\bibfnamefont{Y.}~\bibnamefont{Nishida}},
  \bibinfo{journal}{Phys. Rev. A} \textbf{\bibinfo{volume}{98}},
  \bibinfo{pages}{063634} (\bibinfo{year}{2018}),
  \urlprefix\url{https://link.aps.org/doi/10.1103/PhysRevA.98.063634}.

\bibitem[{\citenamefont{Nishida}(2019)}]{Nishida19}
\bibinfo{author}{\bibfnamefont{Y.}~\bibnamefont{Nishida}},
  \bibinfo{journal}{Annals of Physics} \textbf{\bibinfo{volume}{410}},
  \bibinfo{pages}{167949} (\bibinfo{year}{2019}), ISSN
  \bibinfo{issn}{0003-4916},
  \urlprefix\url{https://www.sciencedirect.com/science/article/pii/S0003491619302040}.

\bibitem[{\citenamefont{Enss}(2019)}]{Enss19}
\bibinfo{author}{\bibfnamefont{T.}~\bibnamefont{Enss}}, \bibinfo{journal}{Phys.
  Rev. Lett.} \textbf{\bibinfo{volume}{123}}, \bibinfo{pages}{205301}
  (\bibinfo{year}{2019}),
  \urlprefix\url{https://link.aps.org/doi/10.1103/PhysRevLett.123.205301}.

\bibitem[{\citenamefont{Hofmann}(2020)}]{Hofmann20}
\bibinfo{author}{\bibfnamefont{J.}~\bibnamefont{Hofmann}},
  \bibinfo{journal}{Phys. Rev. A} \textbf{\bibinfo{volume}{101}},
  \bibinfo{pages}{013620} (\bibinfo{year}{2020}),
  \urlprefix\url{https://link.aps.org/doi/10.1103/PhysRevA.101.013620}.

\bibitem[{\citenamefont{Fujii and Enss}(2024)}]{Fujii24}
\bibinfo{author}{\bibfnamefont{K.}~\bibnamefont{Fujii}} \bibnamefont{and}
  \bibinfo{author}{\bibfnamefont{T.}~\bibnamefont{Enss}},
  \bibinfo{journal}{Phys. Rev. Lett.} \textbf{\bibinfo{volume}{133}},
  \bibinfo{pages}{173402} (\bibinfo{year}{2024}),
  \urlprefix\url{https://link.aps.org/doi/10.1103/PhysRevLett.133.173402}.

\bibitem[{\citenamefont{Mazeliauskas and Enss}(2026)}]{Mazeliauskas26}
\bibinfo{author}{\bibfnamefont{A.}~\bibnamefont{Mazeliauskas}}
  \bibnamefont{and} \bibinfo{author}{\bibfnamefont{T.}~\bibnamefont{Enss}},
  \bibinfo{journal}{Phys. Rev. Lett.} \textbf{\bibinfo{volume}{136}},
  \bibinfo{pages}{103402} (\bibinfo{year}{2026}),
  \urlprefix\url{https://link.aps.org/doi/10.1103/mtgf-f4f3}.

\bibitem[{\citenamefont{Son and Thompson}(2010)}]{Son10}
\bibinfo{author}{\bibfnamefont{D.~T.} \bibnamefont{Son}} \bibnamefont{and}
  \bibinfo{author}{\bibfnamefont{E.~G.} \bibnamefont{Thompson}},
  \bibinfo{journal}{Phys. Rev. A} \textbf{\bibinfo{volume}{81}},
  \bibinfo{pages}{063634} (\bibinfo{year}{2010}),
  \urlprefix\url{https://link.aps.org/doi/10.1103/PhysRevA.81.063634}.

\bibitem[{\citenamefont{Huang and Cazalilla}(2019)}]{Huang19}
\bibinfo{author}{\bibfnamefont{C.-H.} \bibnamefont{Huang}} \bibnamefont{and}
  \bibinfo{author}{\bibfnamefont{M.~A.} \bibnamefont{Cazalilla}},
  \bibinfo{journal}{Phys. Rev. A} \textbf{\bibinfo{volume}{99}},
  \bibinfo{pages}{063612} (\bibinfo{year}{2019}),
  \urlprefix\url{https://link.aps.org/doi/10.1103/PhysRevA.99.063612}.

\bibitem[{\citenamefont{Natu and Mueller}(2013)}]{Natu13}
\bibinfo{author}{\bibfnamefont{S.~S.} \bibnamefont{Natu}} \bibnamefont{and}
  \bibinfo{author}{\bibfnamefont{E.~J.} \bibnamefont{Mueller}},
  \bibinfo{journal}{Phys. Rev. A} \textbf{\bibinfo{volume}{87}},
  \bibinfo{pages}{053607} (\bibinfo{year}{2013}),
  \urlprefix\url{https://link.aps.org/doi/10.1103/PhysRevA.87.053607}.

\bibitem[{\citenamefont{Corson and Bohn}(2015)}]{Corson15}
\bibinfo{author}{\bibfnamefont{J.~P.} \bibnamefont{Corson}} \bibnamefont{and}
  \bibinfo{author}{\bibfnamefont{J.~L.} \bibnamefont{Bohn}},
  \bibinfo{journal}{Phys. Rev. A} \textbf{\bibinfo{volume}{91}},
  \bibinfo{pages}{013616} (\bibinfo{year}{2015}),
  \urlprefix\url{https://link.aps.org/doi/10.1103/PhysRevA.91.013616}.

\bibitem[{\citenamefont{Mu\~noz de~las Heras et~al.}(2019)\citenamefont{Mu\~noz
  de~las Heras, Parish, and Marchetti}}]{Heras19}
\bibinfo{author}{\bibfnamefont{A.}~\bibnamefont{Mu\~noz de~las Heras}},
  \bibinfo{author}{\bibfnamefont{M.~M.} \bibnamefont{Parish}},
  \bibnamefont{and} \bibinfo{author}{\bibfnamefont{F.~M.}
  \bibnamefont{Marchetti}}, \bibinfo{journal}{Phys. Rev. A}
  \textbf{\bibinfo{volume}{99}}, \bibinfo{pages}{023623}
  (\bibinfo{year}{2019}),
  \urlprefix\url{https://link.aps.org/doi/10.1103/PhysRevA.99.023623}.

\bibitem[{\citenamefont{Xie et~al.}(2026)\citenamefont{Xie, Dale, Grehan, Wang,
  Enss, Julienne, Yu, and Thywissen}}]{Xie2026}
\bibinfo{author}{\bibfnamefont{K.~G.~S.} \bibnamefont{Xie}},
  \bibinfo{author}{\bibfnamefont{C.~J.} \bibnamefont{Dale}},
  \bibinfo{author}{\bibfnamefont{K.~P.} \bibnamefont{Grehan}},
  \bibinfo{author}{\bibfnamefont{M.~F.} \bibnamefont{Wang}},
  \bibinfo{author}{\bibfnamefont{T.}~\bibnamefont{Enss}},
  \bibinfo{author}{\bibfnamefont{P.~S.} \bibnamefont{Julienne}},
  \bibinfo{author}{\bibfnamefont{Z.}~\bibnamefont{Yu}}, \bibnamefont{and}
  \bibinfo{author}{\bibfnamefont{J.~H.} \bibnamefont{Thywissen}},
  \bibinfo{journal}{Phys. Rev. Lett.} \textbf{\bibinfo{volume}{136}},
  \bibinfo{pages}{083402} (\bibinfo{year}{2026}),
  \urlprefix\url{https://link.aps.org/doi/10.1103/44rc-34n1}.

\bibitem[{\citenamefont{Partridge et~al.}(2005)\citenamefont{Partridge,
  Strecker, Kamar, Jack, and Hulet}}]{Partridge05}
\bibinfo{author}{\bibfnamefont{G.~B.} \bibnamefont{Partridge}},
  \bibinfo{author}{\bibfnamefont{K.~E.} \bibnamefont{Strecker}},
  \bibinfo{author}{\bibfnamefont{R.~I.} \bibnamefont{Kamar}},
  \bibinfo{author}{\bibfnamefont{M.~W.} \bibnamefont{Jack}}, \bibnamefont{and}
  \bibinfo{author}{\bibfnamefont{R.~G.} \bibnamefont{Hulet}},
  \bibinfo{journal}{Phys. Rev. Lett.} \textbf{\bibinfo{volume}{95}},
  \bibinfo{pages}{020404} (\bibinfo{year}{2005}),
  \urlprefix\url{https://link.aps.org/doi/10.1103/PhysRevLett.95.020404}.

\bibitem[{\citenamefont{Werner et~al.}(2009)\citenamefont{Werner, Tarruell, and
  Castin}}]{Werner09}
\bibinfo{author}{\bibfnamefont{F.}~\bibnamefont{Werner}},
  \bibinfo{author}{\bibfnamefont{L.}~\bibnamefont{Tarruell}}, \bibnamefont{and}
  \bibinfo{author}{\bibfnamefont{Y.}~\bibnamefont{Castin}},
  \bibinfo{journal}{The European Physical Journal B}
  \textbf{\bibinfo{volume}{68}}, \bibinfo{pages}{401} (\bibinfo{year}{2009}),
  ISSN \bibinfo{issn}{1434-6036},
  \urlprefix\url{https://doi.org/10.1140/epjb/e2009-00040-8}.

\bibitem[{\citenamefont{Wang et~al.}(2021)\citenamefont{Wang, Liu, and
  Hu}}]{Wang21}
\bibinfo{author}{\bibfnamefont{J.}~\bibnamefont{Wang}},
  \bibinfo{author}{\bibfnamefont{X.-J.} \bibnamefont{Liu}}, \bibnamefont{and}
  \bibinfo{author}{\bibfnamefont{H.}~\bibnamefont{Hu}}, \bibinfo{journal}{Phys.
  Rev. A} \textbf{\bibinfo{volume}{104}}, \bibinfo{pages}{063309}
  (\bibinfo{year}{2021}),
  \urlprefix\url{https://link.aps.org/doi/10.1103/PhysRevA.104.063309}.

\bibitem[{\citenamefont{Paintner et~al.}(2019)\citenamefont{Paintner, Hoffmann,
  J\"ager, Limmer, Schoch, Deissler, Pini, Pieri, Calvanese~Strinati, Chin
  et~al.}}]{Paintner19}
\bibinfo{author}{\bibfnamefont{T.}~\bibnamefont{Paintner}},
  \bibinfo{author}{\bibfnamefont{D.~K.} \bibnamefont{Hoffmann}},
  \bibinfo{author}{\bibfnamefont{M.}~\bibnamefont{J\"ager}},
  \bibinfo{author}{\bibfnamefont{W.}~\bibnamefont{Limmer}},
  \bibinfo{author}{\bibfnamefont{W.}~\bibnamefont{Schoch}},
  \bibinfo{author}{\bibfnamefont{B.}~\bibnamefont{Deissler}},
  \bibinfo{author}{\bibfnamefont{M.}~\bibnamefont{Pini}},
  \bibinfo{author}{\bibfnamefont{P.}~\bibnamefont{Pieri}},
  \bibinfo{author}{\bibfnamefont{G.}~\bibnamefont{Calvanese~Strinati}},
  \bibinfo{author}{\bibfnamefont{C.}~\bibnamefont{Chin}}, \bibnamefont{et~al.},
  \bibinfo{journal}{Phys. Rev. A} \textbf{\bibinfo{volume}{99}},
  \bibinfo{pages}{053617} (\bibinfo{year}{2019}),
  \urlprefix\url{https://link.aps.org/doi/10.1103/PhysRevA.99.053617}.

\bibitem[{\citenamefont{Liu et~al.}(2022)\citenamefont{Liu, Yao, Chen, Wang,
  Wang, Chen, Chen, Levin, and Pan}}]{Liu22}
\bibinfo{author}{\bibfnamefont{X.-P.} \bibnamefont{Liu}},
  \bibinfo{author}{\bibfnamefont{X.-C.} \bibnamefont{Yao}},
  \bibinfo{author}{\bibfnamefont{H.-Z.} \bibnamefont{Chen}},
  \bibinfo{author}{\bibfnamefont{X.-Q.} \bibnamefont{Wang}},
  \bibinfo{author}{\bibfnamefont{Y.-X.} \bibnamefont{Wang}},
  \bibinfo{author}{\bibfnamefont{Y.-A.} \bibnamefont{Chen}},
  \bibinfo{author}{\bibfnamefont{Q.}~\bibnamefont{Chen}},
  \bibinfo{author}{\bibfnamefont{K.}~\bibnamefont{Levin}}, \bibnamefont{and}
  \bibinfo{author}{\bibfnamefont{J.-W.} \bibnamefont{Pan}},
  \bibinfo{journal}{National Science Review} \textbf{\bibinfo{volume}{9}},
  \bibinfo{pages}{nwab226} (\bibinfo{year}{2022}), ISSN
  \bibinfo{issn}{2095-5138},
  \urlprefix\url{https://doi.org/10.1093/nsr/nwab226}.

\bibitem[{\citenamefont{J\"ager and Denschlag}(2024{\natexlab{a}})}]{Jager24_1}
\bibinfo{author}{\bibfnamefont{M.}~\bibnamefont{J\"ager}} \bibnamefont{and}
  \bibinfo{author}{\bibfnamefont{J.~H.} \bibnamefont{Denschlag}},
  \bibinfo{journal}{Phys. Rev. Lett.} \textbf{\bibinfo{volume}{132}},
  \bibinfo{pages}{263401} (\bibinfo{year}{2024}{\natexlab{a}}),
  \urlprefix\url{https://link.aps.org/doi/10.1103/PhysRevLett.132.263401}.

\bibitem[{\citenamefont{J\"ager and Denschlag}(2024{\natexlab{b}})}]{Jager24_2}
\bibinfo{author}{\bibfnamefont{M.}~\bibnamefont{J\"ager}} \bibnamefont{and}
  \bibinfo{author}{\bibfnamefont{J.~H.} \bibnamefont{Denschlag}},
  \bibinfo{journal}{Phys. Rev. A} \textbf{\bibinfo{volume}{109}},
  \bibinfo{pages}{063330} (\bibinfo{year}{2024}{\natexlab{b}}),
  \urlprefix\url{https://link.aps.org/doi/10.1103/PhysRevA.109.063330}.

\bibitem[{\citenamefont{Journeaux et~al.}(2026)\citenamefont{Journeaux,
  Veschambre, Lecomte, Uzan, Dalibard, Werner, Petrov, and
  Lopes}}]{Journeaux26}
\bibinfo{author}{\bibfnamefont{A.}~\bibnamefont{Journeaux}},
  \bibinfo{author}{\bibfnamefont{J.}~\bibnamefont{Veschambre}},
  \bibinfo{author}{\bibfnamefont{M.}~\bibnamefont{Lecomte}},
  \bibinfo{author}{\bibfnamefont{E.}~\bibnamefont{Uzan}},
  \bibinfo{author}{\bibfnamefont{J.}~\bibnamefont{Dalibard}},
  \bibinfo{author}{\bibfnamefont{F.}~\bibnamefont{Werner}},
  \bibinfo{author}{\bibfnamefont{D.~S.} \bibnamefont{Petrov}},
  \bibnamefont{and} \bibinfo{author}{\bibfnamefont{R.}~\bibnamefont{Lopes}},
  \bibinfo{journal}{Phys. Rev. Lett.} \textbf{\bibinfo{volume}{136}},
  \bibinfo{pages}{083404} (\bibinfo{year}{2026}),
  \urlprefix\url{https://link.aps.org/doi/10.1103/prf8-3q27}.

\bibitem[{\citenamefont{Yang}(1962)}]{Yang1962}
\bibinfo{author}{\bibfnamefont{C.~N.} \bibnamefont{Yang}},
  \bibinfo{journal}{Reviews of Modern Physics} \textbf{\bibinfo{volume}{34}},
  \bibinfo{pages}{694} (\bibinfo{year}{1962}).

\bibitem[{\citenamefont{Josephson}(1962)}]{Josephson1962}
\bibinfo{author}{\bibfnamefont{B.}~\bibnamefont{Josephson}},
  \bibinfo{journal}{Physics Letters} \textbf{\bibinfo{volume}{1}},
  \bibinfo{pages}{251} (\bibinfo{year}{1962}), ISSN \bibinfo{issn}{0031-9163},
  \urlprefix\url{https://www.sciencedirect.com/science/article/pii/0031916362913690}.

\bibitem[{\citenamefont{Petrov et~al.}(2004)\citenamefont{Petrov, Salomon, and
  Shlyapnikov}}]{Petrov2004}
\bibinfo{author}{\bibfnamefont{D.~S.} \bibnamefont{Petrov}},
  \bibinfo{author}{\bibfnamefont{C.}~\bibnamefont{Salomon}}, \bibnamefont{and}
  \bibinfo{author}{\bibfnamefont{G.~V.} \bibnamefont{Shlyapnikov}},
  \bibinfo{journal}{Phys. Rev. Lett.} \textbf{\bibinfo{volume}{93}},
  \bibinfo{pages}{090404} (\bibinfo{year}{2004}).

\bibitem[{\citenamefont{Braaten and Hammer}(2006)}]{BraatenHammer2006}
\bibinfo{author}{\bibfnamefont{E.}~\bibnamefont{Braaten}} \bibnamefont{and}
  \bibinfo{author}{\bibfnamefont{H.-W.} \bibnamefont{Hammer}},
  \bibinfo{journal}{Physics Reports} \textbf{\bibinfo{volume}{428}},
  \bibinfo{pages}{259} (\bibinfo{year}{2006}).

\bibitem[{\citenamefont{Bardeen et~al.}(1957)\citenamefont{Bardeen, Cooper, and
  Schrieffer}}]{BCS1957}
\bibinfo{author}{\bibfnamefont{J.}~\bibnamefont{Bardeen}},
  \bibinfo{author}{\bibfnamefont{L.~N.} \bibnamefont{Cooper}},
  \bibnamefont{and} \bibinfo{author}{\bibfnamefont{J.~R.}
  \bibnamefont{Schrieffer}}, \bibinfo{journal}{Phys. Rev.}
  \textbf{\bibinfo{volume}{108}}, \bibinfo{pages}{1175} (\bibinfo{year}{1957}).

\bibitem[{\citenamefont{Leggett}(1980)}]{Leggett1980}
\bibinfo{author}{\bibfnamefont{A.~J.} \bibnamefont{Leggett}}, in
  \emph{\bibinfo{booktitle}{Modern Trends in the Theory of Condensed Matter}},
  edited by \bibinfo{editor}{\bibfnamefont{A.}~\bibnamefont{Pekalski}}
  \bibnamefont{and} \bibinfo{editor}{\bibfnamefont{J.}~\bibnamefont{Przystawa}}
  (\bibinfo{publisher}{Springer}, \bibinfo{address}{Berlin},
  \bibinfo{year}{1980}), vol. \bibinfo{volume}{115} of
  \emph{\bibinfo{series}{Lecture Notes in Physics}}, pp.
  \bibinfo{pages}{13--27}.

\bibitem[{\citenamefont{Giorgini et~al.}(2008)\citenamefont{Giorgini,
  Pitaevskii, and Stringari}}]{Giorgini2008}
\bibinfo{author}{\bibfnamefont{S.}~\bibnamefont{Giorgini}},
  \bibinfo{author}{\bibfnamefont{L.~P.} \bibnamefont{Pitaevskii}},
  \bibnamefont{and}
  \bibinfo{author}{\bibfnamefont{S.}~\bibnamefont{Stringari}},
  \bibinfo{journal}{Rev. Mod. Phys.} \textbf{\bibinfo{volume}{80}},
  \bibinfo{pages}{1215} (\bibinfo{year}{2008}).

\bibitem[{\citenamefont{Haussmann et~al.}(2009)\citenamefont{Haussmann, Punk,
  and Zwerger}}]{Haussmann2009}
\bibinfo{author}{\bibfnamefont{R.}~\bibnamefont{Haussmann}},
  \bibinfo{author}{\bibfnamefont{M.}~\bibnamefont{Punk}}, \bibnamefont{and}
  \bibinfo{author}{\bibfnamefont{W.}~\bibnamefont{Zwerger}},
  \bibinfo{journal}{Phys. Rev. A} \textbf{\bibinfo{volume}{80}},
  \bibinfo{pages}{063612} (\bibinfo{year}{2009}).

\bibitem[{SM()}]{SM}
\bibinfo{note}{See the Supplementary Material for a detailed derivation within
  the BCS mean-field theory.}

\bibitem[{\citenamefont{Liu and Hu}(2005)}]{LiuHu2005}
\bibinfo{author}{\bibfnamefont{X.-J.} \bibnamefont{Liu}} \bibnamefont{and}
  \bibinfo{author}{\bibfnamefont{H.}~\bibnamefont{Hu}}, \bibinfo{journal}{Phys.
  Rev. A} \textbf{\bibinfo{volume}{72}}, \bibinfo{pages}{063613}
  (\bibinfo{year}{2005}).

\bibitem[{\citenamefont{Klimin et~al.}(2019)\citenamefont{Klimin, Tempere, and
  Kurkjian}}]{KliminTempereKurkjian2019}
\bibinfo{author}{\bibfnamefont{S.~N.} \bibnamefont{Klimin}},
  \bibinfo{author}{\bibfnamefont{J.}~\bibnamefont{Tempere}}, \bibnamefont{and}
  \bibinfo{author}{\bibfnamefont{H.}~\bibnamefont{Kurkjian}},
  \bibinfo{journal}{Phys. Rev. A} \textbf{\bibinfo{volume}{100}},
  \bibinfo{pages}{063634} (\bibinfo{year}{2019}).

\bibitem[{\citenamefont{{Van Loon} et~al.}(2020)\citenamefont{{Van Loon},
  Tempere, and Kurkjian}}]{VanLoonTempereKurkjian2020}
\bibinfo{author}{\bibfnamefont{S.}~\bibnamefont{{Van Loon}}},
  \bibinfo{author}{\bibfnamefont{J.}~\bibnamefont{Tempere}}, \bibnamefont{and}
  \bibinfo{author}{\bibfnamefont{H.}~\bibnamefont{Kurkjian}},
  \bibinfo{journal}{Phys. Rev. Lett.} \textbf{\bibinfo{volume}{124}},
  \bibinfo{pages}{073404} (\bibinfo{year}{2020}).

\bibitem[{\citenamefont{Tsuchiya et~al.}(2009)\citenamefont{Tsuchiya, Watanabe,
  and Ohashi}}]{TsuchiyaWatanabeOhashi2009}
\bibinfo{author}{\bibfnamefont{S.}~\bibnamefont{Tsuchiya}},
  \bibinfo{author}{\bibfnamefont{R.}~\bibnamefont{Watanabe}}, \bibnamefont{and}
  \bibinfo{author}{\bibfnamefont{Y.}~\bibnamefont{Ohashi}},
  \bibinfo{journal}{Phys. Rev. A} \textbf{\bibinfo{volume}{80}},
  \bibinfo{pages}{033613} (\bibinfo{year}{2009}).

\bibitem[{\citenamefont{Gaebler et~al.}(2010)\citenamefont{Gaebler, Stewart,
  Drake, Jin, Perali, Pieri, and Strinati}}]{GaeblerEtAl2010}
\bibinfo{author}{\bibfnamefont{J.~P.} \bibnamefont{Gaebler}},
  \bibinfo{author}{\bibfnamefont{J.~T.} \bibnamefont{Stewart}},
  \bibinfo{author}{\bibfnamefont{T.~E.} \bibnamefont{Drake}},
  \bibinfo{author}{\bibfnamefont{D.~S.} \bibnamefont{Jin}},
  \bibinfo{author}{\bibfnamefont{A.}~\bibnamefont{Perali}},
  \bibinfo{author}{\bibfnamefont{P.}~\bibnamefont{Pieri}}, \bibnamefont{and}
  \bibinfo{author}{\bibfnamefont{G.~C.} \bibnamefont{Strinati}},
  \bibinfo{journal}{Nature Physics} \textbf{\bibinfo{volume}{6}},
  \bibinfo{pages}{569} (\bibinfo{year}{2010}).

\bibitem[{\citenamefont{Combescot et~al.}(2006)\citenamefont{Combescot, Kagan,
  and Stringari}}]{CombescotKaganStringari2006}
\bibinfo{author}{\bibfnamefont{R.}~\bibnamefont{Combescot}},
  \bibinfo{author}{\bibfnamefont{M.~Y.} \bibnamefont{Kagan}}, \bibnamefont{and}
  \bibinfo{author}{\bibfnamefont{S.}~\bibnamefont{Stringari}},
  \bibinfo{journal}{Phys. Rev. A} \textbf{\bibinfo{volume}{74}},
  \bibinfo{pages}{042717} (\bibinfo{year}{2006}).

\bibitem[{\citenamefont{Kurkjian et~al.}(2019)\citenamefont{Kurkjian, Klimin,
  Tempere, and Castin}}]{KurkjianKliminTempereCastin2019}
\bibinfo{author}{\bibfnamefont{H.}~\bibnamefont{Kurkjian}},
  \bibinfo{author}{\bibfnamefont{S.~N.} \bibnamefont{Klimin}},
  \bibinfo{author}{\bibfnamefont{J.}~\bibnamefont{Tempere}}, \bibnamefont{and}
  \bibinfo{author}{\bibfnamefont{Y.}~\bibnamefont{Castin}},
  \bibinfo{journal}{Phys. Rev. Lett.} \textbf{\bibinfo{volume}{122}},
  \bibinfo{pages}{093403} (\bibinfo{year}{2019}).

\bibitem[{\citenamefont{Perali et~al.}(2008)\citenamefont{Perali, Pieri, and
  Strinati}}]{PeraliPieriStrinati2008}
\bibinfo{author}{\bibfnamefont{A.}~\bibnamefont{Perali}},
  \bibinfo{author}{\bibfnamefont{P.}~\bibnamefont{Pieri}}, \bibnamefont{and}
  \bibinfo{author}{\bibfnamefont{G.~C.} \bibnamefont{Strinati}},
  \bibinfo{journal}{Phys. Rev. Lett.} \textbf{\bibinfo{volume}{100}},
  \bibinfo{pages}{010402} (\bibinfo{year}{2008}).

\bibitem[{\citenamefont{Ramos~Rodr{\'\i}guez
  et~al.}(2026)\citenamefont{Ramos~Rodr{\'\i}guez, Gievers, and
  Schmidt}}]{RamosRodriguez2026}
\bibinfo{author}{\bibfnamefont{E.}~\bibnamefont{Ramos~Rodr{\'\i}guez}},
  \bibinfo{author}{\bibfnamefont{M.}~\bibnamefont{Gievers}}, \bibnamefont{and}
  \bibinfo{author}{\bibfnamefont{R.}~\bibnamefont{Schmidt}},
  \bibinfo{journal}{Phys. Rev. Lett.} \textbf{\bibinfo{volume}{136}},
  \bibinfo{pages}{243401} (\bibinfo{year}{2026}).

\end{thebibliography}
\end{document}